\documentclass[twocolumn,showpacs,preprintnumbers]{revtex4}
\usepackage{amsfonts}
\usepackage{amsmath}
\usepackage{amssymb}
\usepackage{graphicx}
\usepackage{graphics}
\usepackage{dcolumn}
\usepackage{bm}

\begin{document}

\title{Two types of phase diagrams for two-species Bose-Einstein condensates}
\author{Z. B. Li$^{1}$, Y. M. Liu$^{2,3}$, D. X. Yao$^1$, and C. G. Bao$^{1}$%
\thanks{
Corresponding author: C.G.Bao, stsbcg@mail.sysu.edu.cn}}
\affiliation{$^{1}$State Key Laboratory of Optoelectronic Materials and Technologies,
School of Physics, Sun Yat-Sen University, Guangzhou, 510275, P. R. China}
\affiliation{$^{2}$Department of Physics, Shaoguan University, Shaoguan, 512005, P. R.
China}
\affiliation{$^{3}$State Key Laboratory of Theoretical Physics, Institute of Theoretical
Physics, Chinese Academy of Sciences, Beijing, 100190, China}

\begin{abstract}
Under the Thomas-Fermi
approximation, a relatively much simpler analytical solutions of the coupled Gross-Pitaevskii
equations for the two-species BEC have been derived. Additionally, a model for the asymmetric
states has been proposed, and the competition between the symmetric and
asymmetric states has been evaluated. The whole
parameter-space is divided into zones, each supports a specific phase, namely, the
symmetric miscible phase, the symmetric immiscible phase, or the asymmetric
phase. Based on the division the phase-diagrams against any set of
parameters can be plotted. Thereby, the effects of these parameters can be
visualized. There are three critical values in the inter-species interaction $%
V_{AB} $ and one in the ratio of particle numbers $N_{A}/N_{B}$. They govern the transitions
between the phases. Two cases, (i)
the repulsive $V_{AB}$ matches the repulsive $%
V_{A}+V_{B}$, and (ii) the attractive $V_{AB}$ nearly cancels
the effect of the repulsive $V_{A}+V_{B}$
have been particularly taken into account. The former leads to a complete separation of the
two kinds of atoms , while the latter lead to a collapse. Finally, based on an equation
derived in the paper, a convenient experimental approach
is proposed to determine the ratio of particle numbers .
\end{abstract}

\maketitle

*Correspondence to stsbcg@mail.sysu.edu.cn

\section{Introduction}

There is increasing interest in two-species Bose-Einstein condensate (TBEC)
in last two decades since it was first studied theoretically by Ho and Shenoy%
\cite{ho96}, and was achieved experimentally by Myatt, et al.\cite{myat97}.
The TBEC provides an important tool to clarify the inter-species and
intra-species interactions. Besides, molecules in well defined rovibrational
levels with permanent electric-dipole moment can be thereby obtained, and
they are valuable for studying the strongly interacting quantum gases.\cite%
{ospe06,webe08,ni08,bara08}. Experimentally, the Rb-Cs mixture has been
realized via a magnetic trap \cite{ande05} or an optical trap \cite{pilc09}.
The Rb-Yb mixture has been realized via a combined magneto-optical trap \cite%
{nemi09}. The K-Rb, Na-Rb, Sr-Rb and the isotopic mixtures $^{87}$Rb-$^{85}$%
Rb, $^{168}$Yb-$^{174}$Yb, $^{84}$Sr-$^{86}$Sr, and $^{86}$Sr-$^{88}$Sr have
also been realized and studied recently (refer to the references listed in
\cite{wack15}).

There are a number of theoretical literatures dedicated to the TBEC \cite%
{ho96,esry97,pu98,timm98,ao98,chui99,tripp00,ribo02,chui03,luo07,xu09,shi11,gau10,nott15, scha15,inde15,roy15,luo08,polo15}%
. The ground state (g.s.) is found to have three phases, symmetric miscible
phase, symmetric immiscible phase, and asymmetric immiscible phase,
depending on the parameters. There are a number of parameters (the strengths
of the intra- and inter-species interactions, the particle numbers, the
masses of atoms, and those for the trap) affecting the behavior of the
system. To clarify the effects of them is a main topic of study.

This paper is dedicated to obtain the phase-diagrams. How the spatial
configuration of the g.s. is affected by the parameters can be thereby
visualized. For this purpose, effort is made to divide the whole
parameter-space into zones, each supports a specific spatial configuration.
Based on the division, two types of phase-diagrams are plotted. Related
analysis is supported by analytical formalism. The emphasis is placed on the
qualitative aspect. The particle numbers are considered to be large ($\geq
10^{4}$) so that the Thomas-Fermi approximation (TFA) can be applied. The
symmetric states are obtained by solving the coupled Gross-Pitaevskii
equations (CGP) in an analytical way. Whereas the asymmetric states are
obtained by introducing a model. Both repulsive and attractive inter-species
interactions are considered. Two critical cases, namely, (i) the repulsive
inter-species interaction matches the repulsive intra-species interaction
and (ii) the attractive inter-species interaction nearly cancels the
repulsive intra-species interaction, are taken into account.

\bigskip

\section{Symmetric ground state}

There are $N_{A}$ A-atoms with mass $m_{A}$\ and interacting via $%
V_{A}=c_{A}\Sigma _{i<i^{\prime }}\delta (\mathbf{r}_{i}\mathbf{-r}%
_{i^{\prime }})$, and $N_{B}$ B-atoms with $m_{B}$\ and $V_{B}=c_{B}\Sigma
_{j<j^{\prime }}\delta (\mathbf{r}_{j}\mathbf{-r}_{j^{\prime }})$. The A-
and B-atoms are interacting via $V_{AB}=c_{AB}\Sigma _{i<j}\delta (\mathbf{r}%
_{i}\mathbf{-r}_{j})$. Their spin-degrees of freedom are considered as being
frozen. They are trapped by the isotropic parabolic potentials $\frac{1}{2}%
m_{S}\omega _{S}^{2}r^{2}$ ($S=A$ or $B$). In the g.s. $\Psi _{gs}$ the
atoms of the same kind will have the same wave function $\phi _{S}$ which is
most advantageous to binding. Thus,
\begin{equation}
\Psi _{gs}=\Pi _{j=1}^{N_{A}}\phi _{A}(\mathbf{r}_{j})\Pi _{k=1}^{N_{B}}\phi
_{B}(\mathbf{r}_{k})  \label{eq1}
\end{equation}

It is well known that $\Psi _{gs}$ and the trap might not have the same
symmetry. When the interspecies interaction is sufficiently repulsive, the
two kinds of atoms might separate from each other in an asymmetric way.
Nonetheless, we consider first the case that the g.s. is symmetric.

In this case, let $\phi _{A}(\mathbf{r}_{j})\equiv \frac{u(r_{j})}{\sqrt{%
4\pi }r_{j}}$ and $\phi _{B}(\mathbf{r}_{k})\equiv \frac{v(r_{k})}{\sqrt{%
4\pi }r_{k}}$. We introduce a mass $m$ and a frequency $\omega $. $\hbar
\omega $ and $\sqrt{\hbar /(m\omega )}$ are used as units for energy and
length in this paper. We further introduce $\gamma _{s}\equiv \frac{m_{s}}{m}%
(\frac{\omega s}{\omega })^{2}$ and a set of parameters $\alpha _{1}\equiv
N_{A}c_{A}/(4\pi \gamma _{A})$, $\beta _{1}\equiv N_{B}c_{AB}/(4\pi \gamma
_{A})$, $\alpha _{2}\equiv N_{B}c_{B}/(4\pi \gamma _{B})$, $\beta _{2}\equiv
N_{A}c_{AB}/(4\pi \gamma _{B})$. This set is called the weighted strengths
(W-strengths). Under the TFA, the CGP for $u$ and $v$ appear as
\begin{eqnarray}
&&(\frac{r^{2}}{2}+\alpha _{1}\frac{u^{2}}{r^{2}}+\beta _{1}\frac{v^{2}}{%
r^{2}}-\varepsilon _{1})u=0  \label{eq5a} \\
&&(\frac{r^{2}}{2}+\beta _{2}\frac{u^{2}}{r^{2}}+\alpha _{2}\frac{v^{2}}{%
r^{2}}-\varepsilon _{2})v=0  \label{eq5b}
\end{eqnarray}%
where $\beta _{2}$ and $\beta _{1}$ for the inter-species interaction have
the same sign ($+$ or $-$) while $\alpha _{2}$ and $\alpha _{1}$\ for the
intra-species interaction are considered to be positive only. The chemical
potential for the A-atoms (B-atoms) is equal to $\gamma _{A}\varepsilon _{1}$
($\gamma _{B}\varepsilon _{2}$). The normalization $\int u^{2}dr=1$ and $%
\int v^{2}dr=1$ are required. $u\geq 0$ and $v\geq 0$ are safely assumed.

Eqs.(\ref{eq5a},\ref{eq5b}) demonstrate that the combined effects of the
nine dynamical parameters $c_{S}$, $c_{AB}$, $N_{S}$, $m_{S}$, and $\omega
_{S}$ ($S=A$ and $B$) are sufficiently embodied by the four W-strengths.
Thus an approach based on the W-strengths would lead to a simplification.
Based on the W-strengths, we define

\begin{eqnarray}
Y_{1} &\equiv &(\alpha _{2}-\beta _{1})/(2\mathbf{\alpha )} \\
Y_{2} &\equiv &(\alpha _{1}-\beta _{2})/(2\mathbf{\alpha )}  \label{6}
\end{eqnarray}

where $\mathbf{\alpha }\equiv \alpha _{1}\alpha _{2}-\beta _{1}\beta _{2}$. $%
Y_{1}$ and $Y_{2}$ are crucial in constituting the solutions of the CGP as
shown below.

It has been known that the spatial configuration of the symmetric g.s. has
two phases. When each kind of atoms form a core surrounding the center, then
it is named miscible phase denoted as (A,B) (if the B-atoms have a broader
distribution) or (B,A). When a kind of atoms form a core while the other
kind form a shell away from the center, it is named immiscible phase denoted
as [A-B] (if the B-atoms form the shell) or [B-A].. More strictly, the
miscible phase has both $u/r|_{r=0}$ and $v/r|_{r=0}$ being nonzero, while
the immiscible phase has one of them being zero.

The analytical solutions of the symmetric g.s. under the TFA have been
worked out previously \cite{tripp00,polo15}. Nonetheless, in terms of the
W-strengths, the analytical solutions could have a considerably simpler
form. This simplification is valuable because it substantially benefits a
more detailed analysis. Thus, the simplified solutions are given in the
Appendix. Besides, the division of the whole parameter-space into zones is a
crucial point in our paper and is also included in the Appendix. With the
division, various types of phase-diagrams can be plotted.

\bigskip

\section{The phase-diagram against the interspecies interaction and the
particle numbers}

Based on the analytical solutions of the CGP and the division of the
parameter-space which are given in the Appendix, the first type of
phase-diagrams are shown in Fig.1 .

\begin{figure}[tbp]
\centering \resizebox{0.95\columnwidth}{!}{\includegraphics{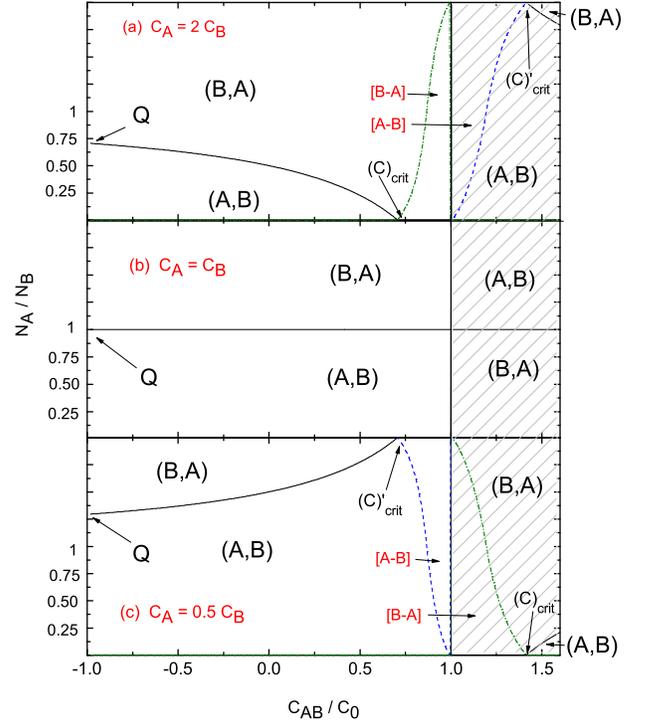} }
\caption{(Color online) The phase diagram with respect to $C_{AB}/C_{0}$ and
$N_{A}/N_{B}$, where $C_{0}\equiv \protect\sqrt{C_{A}C_{B}}$. $\protect%
\gamma _{A}=\protect\gamma _{B}=1$, $N_{B}=10^{4}$, $C_{B}=10^{-3}$, and $%
C_{A}$\ is given at three values marked in the panels. The units $\hbar
\protect\omega $\ and $\protect\sqrt{\hbar /(m\protect\omega )}$ are used in
this paper. The ordinate $y$\ is given from 0 to 2. When $y\leq 1$, $%
N_{A}/N_{B}=y$. When $y>1$, $N_{A}/N_{B}=1/(2-y)$ (in other words, $%
N_{A}/N_{B}$ is ranged from 0 to $\infty $). The labels of phase (e.g.,
(A,B)) are directly marked in the associated zone. The curve separating
(A,B) and (B,A) fulfills $Y_{1}=Y_{2}$, at which $u$ and $v$\ overlap
completely (when $Y_{1}=Y_{2}$, we have $\protect\alpha _{2}-\protect\beta %
_{1}=\protect\alpha _{1}-\protect\beta _{2}$, then one can directly prove
that $u=v$\ is a solution of the CGP). The dash curve separating (A,B) and $%
[A-B]$ fulfills $X_{2}=0$ (i.e., $(1+\frac{\protect\alpha _{2}}{\protect%
\beta _{2}})^{2/3}=2\protect\beta _{2}Y_{1}$, refer to eq.(\protect\ref{eq16}%
)). The dash-dot curve separating (B,A) and $[B-A]$ fulfills $X_{1}^{\prime
}=0$ (i.e., $(1+\frac{\protect\alpha _{1}}{\protect\beta _{1}})^{2/3}=2%
\protect\beta _{1}Y_{2}$).\ The shaded area has $C_{AB}>C_{0}$ (or $\mathbf{%
\protect\alpha }<0$) where the g.s. is asymmetric while the symmetric
solution of the CGP is a collectively excited metastable state. The point $Q$
marks the limit of $(N_{A}/N_{B})_{crit}$ when $C_{AB} \rightarrow -C_{0}$.}
\end{figure}

Let us define a critical value $(C)_{crit}\equiv \frac{\gamma _{A}}{\gamma
_{B}}\sqrt{\frac{C_{B}}{C_{A}}}$. In Fig.1 the lower and upper ends of the
dash-dot curve have $C_{AB}/C_{0}=(C)_{crit}$ and $1$, respectively. The two
ends of the dash curve have $C_{AB}/C_{0}=1$ and $(C)_{crit}^{\prime }\equiv
1/(C)_{crit}$, respectively. Accordingly, $C_{AB}/C_{0}$ contains three
critical values $(C)_{crit}$, 1, and $(C)_{crit}^{\prime }$. In Fig.1a, when
$C_{AB}/C_{0}\leq -1$, the system collapses (see below). When $%
-1<C_{AB}/C_{0}\leq (C)_{crit}$, the g.s. is in miscible phase. There is a
competition between (A,B) and (B,A) phases depending on $N_{A}/N_{B}$. When $%
N_{A}$\ is sufficiently small so that $Y_{1}\geq Y_{2}$, then the
distribution of the A-atoms will be narrower and the g.s. will be in (A,B)
(refer to the Appendix), or vice versa. Thus, from the condition $Y_{1}=Y_{2}
$, we can define a critical ratio
\begin{equation}
(N_{A}/N_{B})_{crit}=(C_{B}\gamma _{A}-C_{AB}\gamma _{B})/(C_{A}\gamma
_{B}-C_{AB}\gamma _{A})\ \ \   \label{7}
\end{equation}

When $N_{A}/N_{B}<$ $(N_{A}/N_{B})_{crit}$, the g.s. is in (A,B), or
otherwise in (B,A). When $N_{A}/N_{B}=(N_{A}/N_{B})_{crit}$, the
distributions of the A- and B-atoms overlap completely (refer to the
Appendix). When $C_{AB}/C_{0}\rightarrow -1$, $(N_{A}/N_{B})_{crit}%
\rightarrow \sqrt{C_{B}/C_{A}}$, thus the condition for the g.s. in (A,B) is
$N_{A}/N_{B}<\sqrt{C_{B}/C_{A}}$ (in other words, the (A,B) phase  would be
more probable if $C_{A}$ is relatively weak and/or $N_{A}$ is small). When $%
C_{AB}/C_{0}=(C)_{crit} $, $(N_{A}/N_{B})_{crit}=0$. When $C_{AB}$ increases
further, the (A,B) phase  disappears and the A-atoms begin to be distributed
more outward. When $(C)_{crit}<C_{AB}/C_{0}\leq 1$, There is a competition
between the [B-A] and (B,A) phases depending also on $N_{A}/N_{B}$. When $%
N_{A}$\ is sufficiently small to enable $X_{1}^{\prime }<0$, the A-atoms
will form a shell and the g.s. is in immiscible phase. Otherwise, in
miscible phase. When $C_{AB}/C_{0}\rightarrow 1$, the zone of [B-A] covers
nearly the whole range of $N_{A}/N_{B}$, and thus the immiscible phase
becomes dominant. When $1<C_{AB}/C_{0}\leq (C)_{crit}^{\prime }$, the lowest
symmetric solutions of the CGP have the B-atoms being distributed more
outward (in [A-B] or (A,B) phases). Nonetheless, when $C_{AB}$\ increases
and crosses $C_{0}$, a transition from [B-A] to [A-B] will occur. During
this transition, the atoms in the shell and in the core interchange their
roles (i.e., those in the core jump to the shell, and vice versa). This
causes a remarkable increase in energy (as shown below). Accordingly, when $%
C_{AB}/C_{0}\geq 1$, the symmetric configuration is no more stable. For this
reason, we shall more concentrate on the case $C_{AB}$ $<C_{0}$.

When $(C)_{crit}=1$, $(C)_{crit}^{\prime }$ also $=1$, and the areas of the
zones of [B-A] and [A-B] both become zero, and therefore the immiscible
phase does not appear (Fig.1b belongs to this case). Thus the difference
between $C_{A}/\gamma _{A}^{2}$ and $C_{B}/\gamma _{B}^{2}$ is a basic
requirement for the appearance of the immiscible phase.

For the case $(C)_{crit}>1$ (Fig.1c belong to this case), a similar
discussion can be performed where the A- and B-atoms interchange their roles.

To conclude, in most cases, the g.s. is in miscible phases. There is a
critical value $(N_{A}/N_{B})_{crit}$ to judge either (A,B) or (B,A) would
win. Besides, there is a critical value $(C)_{crit}$. When $%
(C)_{crit}<C_{AB}/C_{0}<1$ and when $N_{A}$\ is so small that $X_{1}^{\prime
}<0$, the immiscible [B-A] phase will emerge in the zone lying between the
curve $X_{1}^{\prime }=0$ and the vertical line $C_{AB}/C_{0}=1$ (where $%
\mathbf{\alpha }=0$). Note that the [B-A] phase requires thIn this zone $%
Y_{2}>0$ and $Y_{1}<0$ (refer to the Appendix), it implies that the
W-strengths should be appropriately given. The condition for [A-B] can be
similarly deduced. Thus the zones of the immiscible phase appear only by the
two sides of the vertical line $C_{AB}/C_{0}=1$. The area of the zones
depends on how far $(C)_{crit}$ deviates from 1. Recall that $(C)_{crit}$
contains the factor $(\omega _{A}/\omega _{B})^{2}$. Thus the ratio of the
two frequencies is a very sensitive factor to affect the spatial
configuration (say, a small reduction of $\omega _{A}$ will reduce $%
(C)_{crit}$, the zone of [B-A] shown in Fig.1a will be thereby broader, and
thus favors the A-atoms to form a shell).

The spatial configurations, namely, the patterns of the wave functions have
already been plotted in many literatures. To avoid tedious, only two
particular cases are plotted here.

(i) For attractive $V_{AB}$ with $C_{AB}\gtrsim -C_{0}$

In this case both $Y_{1}$ and $Y_{2}$ are positive and tend to $\infty $.
Let us assume $Y_{1}>Y_{2}$. Accordingly, the A-atoms have a narrower
distribution. Since the radius of the inner core $r_{in}\propto Y_{1}^{-1/5}$%
(Refer to the Appendix), it tends to zero. On the other hand, the outer
border $r_{out}=\sqrt{2\varepsilon _{2}}=[15(\alpha _{2}+\beta _{2})]^{1/5}$%
. When $C_{AB}\rightarrow -C_{0}$, $\alpha _{2}+\beta _{2}\rightarrow \frac{%
N_{B}C_{0}}{4\pi \gamma _{B}}(\sqrt{\frac{C_{B}}{C_{A}}}-\frac{N_{A}}{N_{B}}%
) $. Recall that $\sqrt{\frac{C_{B}}{C_{A}}}$ is the limit of $%
(N_{A}/N_{B})_{crit}$ when $C_{AB}\rightarrow -C_{0}$. Thus, if $N_{A}/N_{B}$
is so given that it would considerably smaller than the limit, then $r_{out}$
would be finite while $r_{in}\rightarrow 0$. In this case, we will have a
very dense core together with a thin tail extending outward formed by the
B-atoms. Alternatively, if the W-strengths are so given that $Y_{1}<Y_{2}$,
then the tail is formed by the A-atoms. Whereas, when $\frac{N_{A}}{N_{B}}%
\rightarrow \sqrt{\frac{C_{B}}{C_{A}}}$, the tail disappears and all the
atoms stay inside a very small core. Numerical calculations support this
suggestion as shown in Fig.2, where the tail is long in 2a and 2b, but very
short in 2c.  The extremely high density at the center implies that the
system tends to collapse.

(ii) For repulsive $V_{AB}$ with $C_{AB}\lesssim C_{0}$

When $C_{AB}\rightarrow C_{0}$, the two species tend to separate completely
from each other (refer to the Appendix). Examples for the cases with $%
C_{AB}=0.99C_{0}$ are given in Fig.3 characterized by having a shell
attached to the outward border of a core. In Fig.3 $N_{B}$ remains
unchanged, while the increase of $N_{A}$\ pushes the shell of the B-atoms
more and more outward.

\begin{figure}[tbp]
\centering \resizebox{0.95\columnwidth}{!}{\includegraphics{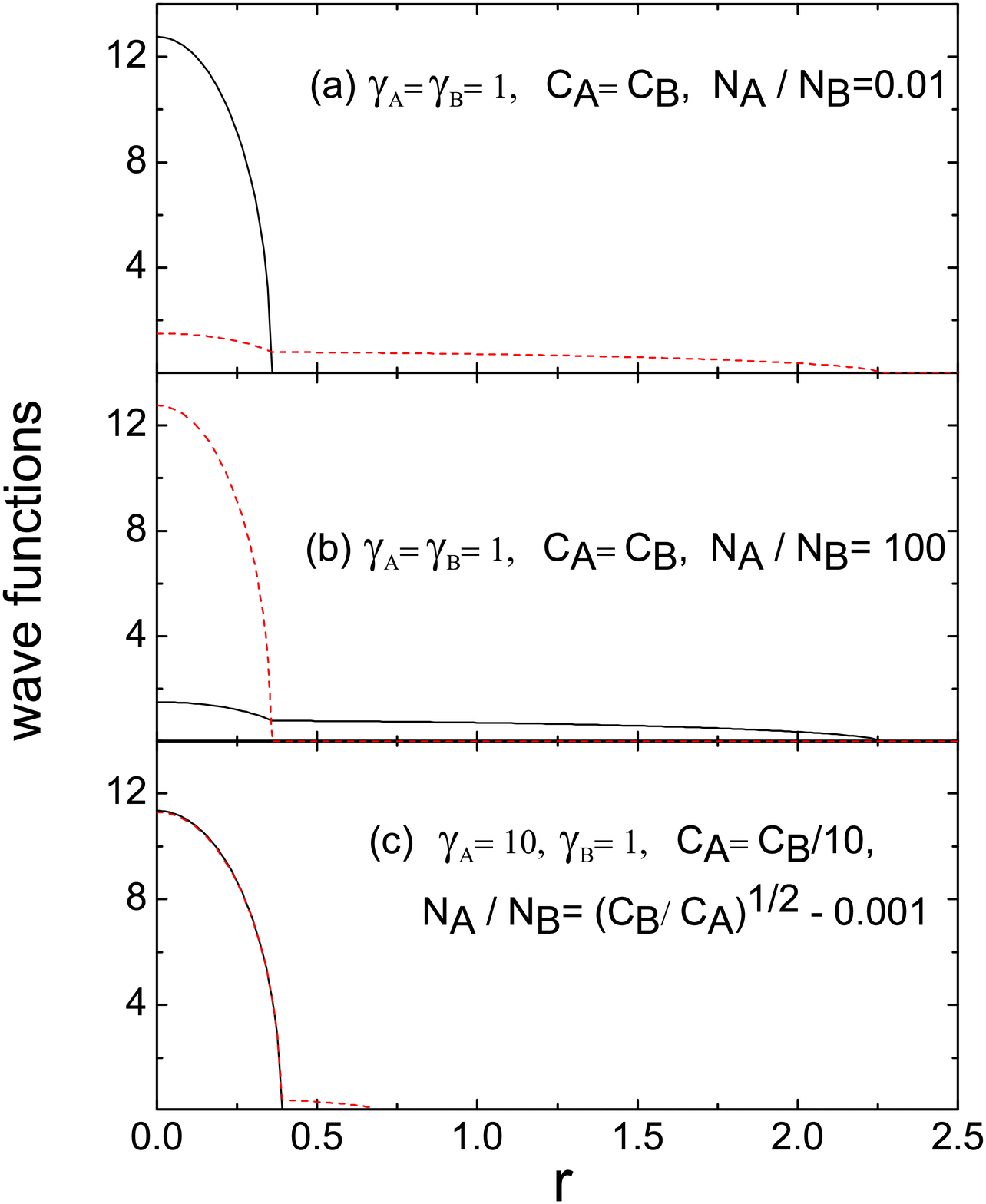} }
\caption{(Color online) $u/r$ (solid) $v/r$\ (dash) against $r$ (in $\protect%
\sqrt{\hbar /(m\protect\omega )}$) when the attraction from $V_{AB}$\ nearly
cancels the repulsion from the intra-species interaction ($C_{AB}=-0.99C_{0}$
is given). $N=N_{A}+N_{B}=5\times 10^{4}$, $C_{B}=10^{-3}$, the other
parameters are marked in the panels. In addition to the high density at the
center, there is a long tail of B-atoms, a long tail of A-atoms, and a very
short tail in a, b, and c.}
\end{figure}

\begin{figure}[tbp]
\centering \resizebox{0.95\columnwidth}{!}{\includegraphics{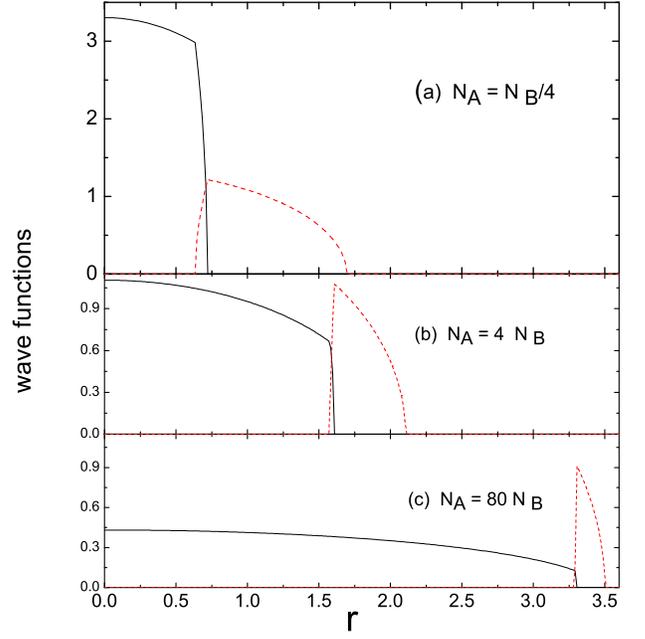} }
\caption{(Color online) $u/r$ (solid) $v/r$\ (dash) against $r$ (in $\protect%
\sqrt{\hbar /(m\protect\omega )}$) when the repulsion from $V_{AB}$\ matches
the repulsion from $V_{A}$ and $V_{B}$ ($C_{AB}=0.99C_{0}$ is given). The
other parameters are $\protect\gamma _{A}=\protect\gamma _{B}=1$, $%
N_{B}=10^{4}$, $C_{B}=10^{-3}$, and $C_{A}=C_{B}/2$. $N_{A}$ is marked in
each panel.}
\end{figure}

\section{Total energy and the competition of symmetric and asymmetric states}

When $\phi _{A}$ and $\phi _{B}$ have been known, the total energy of the
system (neglecting the kinetic energy) is
\begin{eqnarray}
E&=&\frac{N_{A}}{2}\langle r^{2}\rangle _{A}+\frac{N_{B}}{2}\langle
r^{2}\rangle _{B}+\frac{N_{A}^{2}}{2}V_{A}\langle \phi _{A}^{2}\rangle _{A}
\notag \\
&&+\frac{N_{B}^{2}}{2}V_{B}\langle \phi _{B}^{2}\rangle
_{B}+N_{A}N_{B}V_{AB}\langle \phi _{A}^{2}\rangle _{B}  \label{8}
\end{eqnarray}
where $\langle r^{2}\rangle _{A}=\int \phi _{A}^{2}r^{2}d\mathbf{r}$, $%
\langle \phi _{A}^{2}\rangle _{A}=\int \phi _{A}^{4}d\mathbf{r}$, $\langle
\phi _{A}^{2}\rangle _{B}=\langle \phi _{B}^{2}\rangle _{A}=\int \phi
_{A}^{2}\phi _{B}^{2}d\mathbf{r}$, and so on.

Examples of $E/N$ (where $N=N_{A}+N_{B}$) are plotted in Fig.4. Where the
five curves for symmetric states goes up with $C_{AB}$ as expected. In all
cases there is a notable jump when $C_{AB}$ increases and crosses $C_{0}$
(strictly, only the case that $C_{AB}$ tends to $\pm C_{0}$ can be
calculated). This jump is associated with the transition from $[A-B]$ to $%
[B-A] $ mentioned above. This figure demonstrates that the increase of
energy in the jump can be quite large. It implies that very high-lying
collectively excited states might exist. All the horizontal lines are for
asymmetric states, they are obtained via a model shown below.

\begin{figure}[tbp]
\centering \resizebox{0.95\columnwidth}{!}{\includegraphics{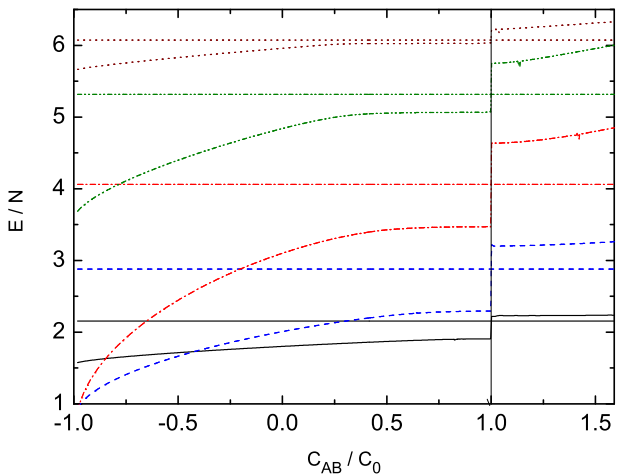} }
\caption{(Color online) $E/N$ (in $\hbar \protect\omega $) of the lowest
symmetric and asymmetric states against $C_{AB}/C_{0}$. The latter are given
by horizontal lines, while the former are noy. The parameters are given as $%
\protect\gamma _{A}=5$, $\protect\gamma _{B}=1$, $N=N_{A}+N_{B}=5\times
10^{4}$, $C_{B}=10^{-3}$, and $C_{A}=2C_{B}$. The solid, dash, dash-dot,
dash-dot-dot, and dotted curves are for $N_{A}/N_{B}=1/20$, 1/4, 1, 4, and
20, respectively.}
\end{figure}

Let $z$\ and $r_{\perp }$ ($z^{2}+r_{\perp }^{2}=r^{2}$) be the axial
coordinates, and two parameters $z_{o}$ and $r_{o}$ are introduced. Then, it
is assumed that when $-r_{o}\leq z\leq z_{o}$
\begin{eqnarray}
u=\frac{1}{c}r\sqrt{1-r^{2}/r_{o}^{2}} \\
v=0  \label{9}
\end{eqnarray}
and when $z_{o}<z\leq r_{o}$
\begin{eqnarray}
u=0 \\
v=\frac{1}{d}r\sqrt{1-r^{2}/r_{o}^{2}}.  \label{10}
\end{eqnarray}
In other words, we have proposed a model in which a ball with radius $r_{o}$
is divided into two sides separated by the plane $z=z_{o}$. The A-atoms and
B-atoms stay separately in the two sides. The orientation of the Z-axis is
irrelevant (because the trap is isotropic). From the normalization, we have
\begin{eqnarray}
c^{2}=\frac{1}{15}r_{o}^{3}+\frac{1}{8}r_{o}^{2}z_{o}-\frac{1}{12}z_{o}^{3}+%
\frac{1}{40}r_{o}^{-2}z_{o}^{5} \\
d^{2}=\frac{1}{15}r_{o}^{3}-\frac{1}{8}r_{o}^{2}z_{o}+\frac{1}{12}z_{o}^{3}-%
\frac{1}{40}r_{o}^{-2}z_{o}^{5}  \label{11}
\end{eqnarray}
The two parameters $r_{o}$\ and $z_{o}$ are determined by minimizing the
total energy $E$ (which is calculated via eq.(\ref{8}) where the
contribution from the kinetic energy has been neglected). The resultant $E/N$%
\ are plotted in Fig.4 and they appear as horizontal lines. In various
cases, due to the jump appearing at $C_{AB}=C_{0}$, the asymmetric states
become lower in energy once $C_{AB}>C_{0}$ (i.e., they become the g.s.).
This is clearly shown in the figure.

Recalled that the kinetic energy has been neglected. When the particle
number is huge, the ratio of the kinetic energy over the total energy is
small (say, in an estimation given in \cite{hlb15}, the ratio is 6\% when $%
N=12000$). Thus the neglect will not spoil the qualitative results.

\section{The phase-diagram against the inter- and intra-species interactions}

The second type of phase-diagrams are given in Fig.5.

\begin{figure}[tbp]
\centering \resizebox{0.95\columnwidth}{!}{\includegraphics{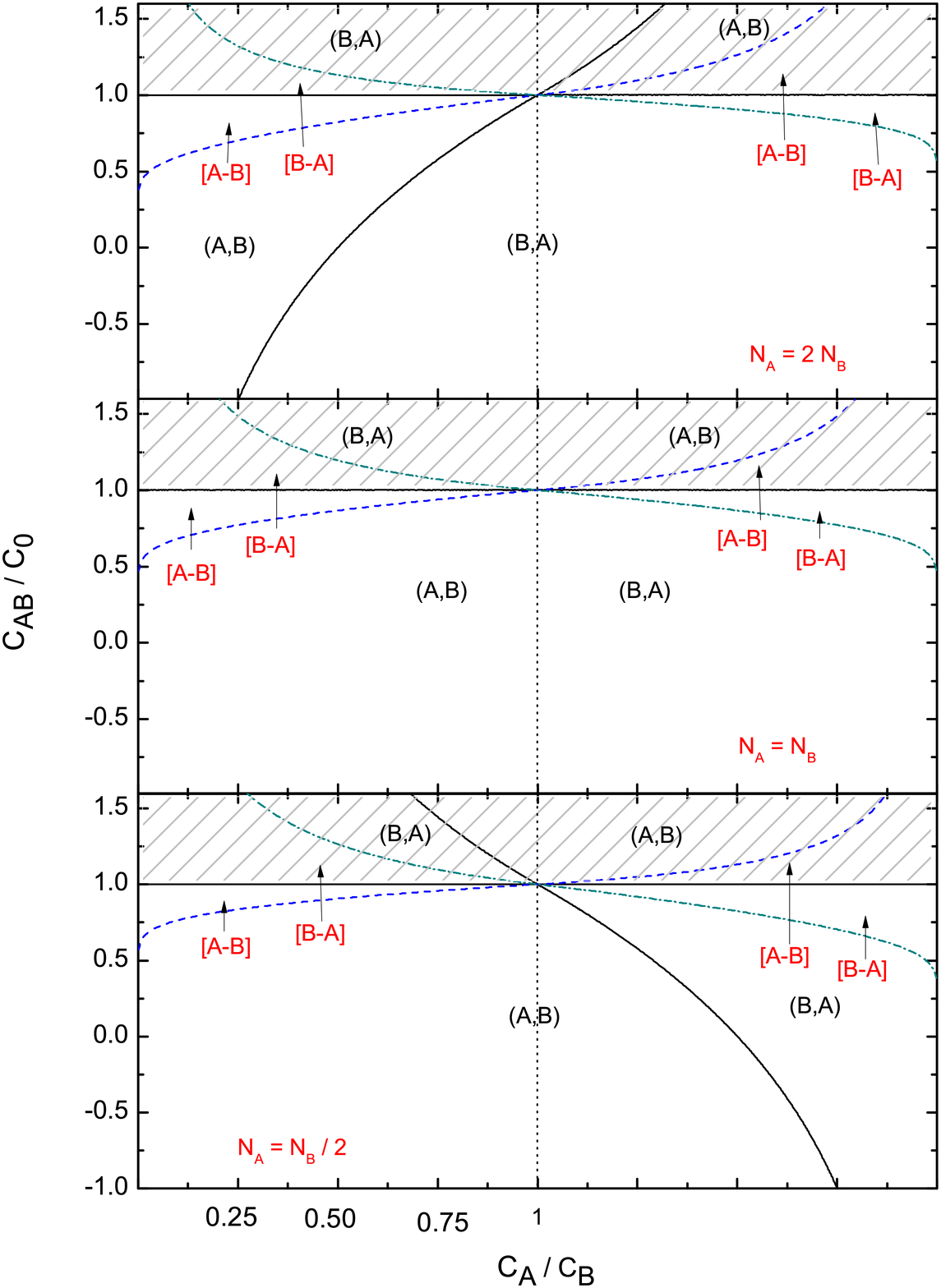} }
\caption{(Color online) The phase diagram with respect to $C_{A}/C_{B}$ and $%
C_{AB}/C_{0}$. $\protect\gamma _{A}=\protect\gamma _{B}=1$, $N_{B}=10^{4}$,
and $N_{A}$ is given at three values marked in the panels. $C_{B}=10^{-3}$.
The abscissa $x$\ for $C_{A}/C_{B}$ is given from 0 to 2. When $x\leq 1$, $%
C_{A}/C_{B}=x$. When $x>1 $, $C_{A}/C_{B}=1/(2-x)$ (in other words, $%
C_{A}/C_{B}$ is ranged from 0 to $\infty $). Refer to Fig.1.}
\end{figure}

The physics extracted from Fig.5 and from Fig.1 are similar. Therefore, we
just mention a few points: (i) The zones for the immiscible phase appear
only in the neighborhood of the horizontal line at which $C_{AB}=C_{0}$.
(ii) For all the cases with $C_{AB}\lesssim C_{0}$, the immiscible and
miscible phases are competing.  when $C_{B}$\ is much stronger than $C_{A}$,
the B-atoms will form a shell and the g.s. is in the [A-B] phase.. When $%
C_{B}$\ is not so strong, the B-atoms will be closer to the core and the
[A-B] becomes (A,B). When $C_{B}$\ reduces further, the B-atoms will have a
narrower distribution while the A-toms will be distributed more outward, and
the (A,B). becomes (B,A). When $C_{B}$ becomes very weak, the A-atoms will
form a shell and the (B,A). becomes [B-A]. (iii) When $C_{AB}$ decreases
further from $C_{0}$, immiscible phase disappears.

\bigskip

\section{Summary and final remarks}

(i) Due to the introduction of the W-strengths, analytical solutions of the
CGP can be expressed in a considerably simpler form to facilitate related analysis.

(ii) A model for the asymmetric g.s. has been proposed. The competition
between the symmetric and asymmetric states has been evaluated. A big gap
in energy was found during the [A-B] to [B-A] (or reversely) transition
which takes place when $C_{AB}$ crosses $C_{0}$. Due to the gap, the
asymmetric states become the g.s. once $C_{AB}>C_{0}$.

It has been known that, in the low-density
region, the overlap of the two wave functions will be remarkably underestimated under
the TFA. Thus the energy $<V_{AB}>$ will be therefore underestimated. This would affect the evaluation of the gap.
However, the gap appears only when a kind of atoms tend to leave completely from the other kind. In this case, the percent
of $<V_{AB}>$ in the total energy is extremely small. Thus, the existence of the gap, which is
the difference of the two total energies, will not be spoiled by using the TFA.

(iii) The whole parameter-space has been divided into zones each supports a
specific spatial configuration. Based on the division two types of
phase-diagrams have been plotted. Making use of the phase-diagrams where the boundaries have analytical forms,
 the effects of various parameters can
be clarified not only qualitatively but also quantitatively.

(iv) Additional effort is made to study the feature of the g.s. when $%
C_{AB}\lesssim C_{0}$ and $C_{AB}\gtrsim -C_{0}$. In the former case the
shell tends to separate from the core completely, while in the latter the
system tends to collapse.

(v) We have found a critical point $(C)_{crit}$ which is also useful in application. One can tune $N_{A}$ and/or
$\omega _{A}$ to make $(C)_{crit}$ tends to $1$ to prohibit the formation of a shell
(i.e.,the shell can not be formed even if
$C_{AB}$ is sufficiently strong), or
to make $(C)_{crit}$ differs remarkably from $1$ to facilitate the formation of a shell (i.e.,the shell can be formed even if
$C_{AB}$ is rather weak).

The point $(C)_{crit}$ is associated with the case that a kind of atoms tend to leave completely
 from the center. In this case TFA is still applicable in the qualitative sense
 (comparing Fig.1b and 1c of \cite{polo15}, where 1b is closer to the above case).

(vi) We have found another critical point $(N_{A}/N_{B})_{crit}$ which marks
the complete overlap of $u/r$ and $v/r$. This finding has potential usage. In experiments, one can tune the tunable parameters
(say, the trap frequency and/or the strengths) so that the two density profiles overlap.
 Then, once all the other parameter are known, $N_{A}/N_{B}$ can be directly known from
eq.(\ref{7}). Since the particle numbers are in general difficult to determine precisely, this approach provides
 an auxiliary way helpful for determining the numbers.

  It was found that in the case that both $u/r$ and $v/r$ are nonzero at $r=0$ and close to each other, the TFA wave functions
deviate very slightly from those beyond the TFA (comparing Fig.1a and 1b of \cite{polo15}).
 Therefore, eq.(\ref{7}) is reliable because it is just associated with this case.

\bigskip

\begin{acknowledgments}
Supported by the National Natural Science Foundation of China under Grants
No.11372122, 11274393, 11574404, and 11275279; the Open Project Program of
State Key Laboratory of Theoretical Physics, Institute of Theoretical
Physics, Chinese Academy of Sciences, China; and the National Basic Research
Program of China (2013CB933601).
\end{acknowledgments}

\section*{Appendix: Analytical solutions of the CGP under TFA and a division
of the parameter-space}

\subsection{Miscible phase}

For the miscible phase (A,B), the normalized solutions $u$ and $v$ are
distributed in two domains of $r$. In the first domain $(0\leq r\leq (\frac{%
15}{2Y_{1}})^{1/5}\equiv r_{a})$ they appear as

\begin{eqnarray}
u^{2}/r^{2} &=&X_{1}-Y_{1}r^{2} \\
v^{2}/r^{2} &=&X_{2}-Y_{2}r^{2}  \label{15}
\end{eqnarray}%
where $Y_{1}$ and $Y_{2}$ are defined in eq.(\ref{6}) depending on the
W-strengths directly,
\begin{eqnarray}
X_{1} &=&(15/2)^{2/5}Y_{1}^{3/5} \\
\varepsilon _{2} &=&\frac{1}{2}[15(\alpha _{2}+\beta _{2})]^{2/5} \\
X_{2} &=&(\varepsilon _{2}-\beta _{2}X_{1})/\alpha _{2}  \label{eq16}
\end{eqnarray}%
Furthermore, when $X_{1}$ and $X_{2}$ are known, $\varepsilon _{1}$ is
related to them as $\varepsilon _{1}=\alpha _{1}X_{1}+\beta _{1}X_{2}$.

In the second domain $(r_{a}<r\leq \sqrt{2\varepsilon _{2}})$%
\begin{eqnarray}
u^{2}/r^{2} &=&0 \\
v^{2}/r^{2} &=&\frac{1}{\alpha _{2}}(\varepsilon _{2}-r^{2}/2)  \label{eq17}
\end{eqnarray}

when $r>\sqrt{2\varepsilon _{2}}$, both $u$ and $v$ are zero. Thus $r_{a}$
and $\sqrt{2\varepsilon _{2}}$ mark the outward borders of the A-atoms and
B-atoms, respectively. One can check directly that the above $u$ and $v$
satisfy the CGP, they are normalized, and they are continuous at the common
border of the domains (however their derivatives are not).

Obviously, the above solution would be physically meaningful only if the
W-strengths are so preset that $Y_{1}>0$, $\alpha _{2}+\beta _{2}>0$, and $%
\sqrt{2\varepsilon _{2}}>r_{a}$. The latter can be satisfied when $Y_{1}\geq
Y_{2}$ is preset. Besides, in order to have both $u/r$ and $v/r$ being $\geq
0$ at $r=0$, $X_{2}\geq 0$ (or $\varepsilon _{2}>\beta _{2}X_{1}$,
equivalently $[2(\alpha _{2}+\beta _{2})Y_{1}]^{2/5}\geq 2\beta _{2}Y_{1}$)
is required. Due to this constraint, the phase (A,B) could emerge only in a
specific zone inside the whole parameter-space. This zone is bound by the
surfaces $Y_{1}=Y_{2}$ and $X_{2}=0$. Inside this zone $Y_{1}>0$, $Y_{1}\geq
Y_{2}$, and $X_{2}>0$\ hold. Referred to Fig.1 and 5.

For attractive $V_{AB}$, $X_{2}>0$ holds always. Thus the border $X_{2}=0$
does not appear. Accordingly, the zone for (A,B) is bound by the surfaces $%
Y_{1}=Y_{2}$ and $C_{AB}=-C_{0}$ (where $C_{0}\equiv \sqrt{C_{A}C_{B}}$).
However, when $C_{AB}\rightarrow -C_{0}$ or $\mathbf{\alpha }\rightarrow 0$,
both $Y_{1}$ and $Y_{2}$ tend to $+\infty $. Hence, $r_{a}$ becomes
extremely small and the system is highly dense at the center. This will lead
to a collapse. Due to the collapse, the case $C_{AB}\leq -C_{0}$ is not
considered.

For (B,A), the solution can be similarly obtained from the above formulae by
an interchange of $u$ and $v$ together with an interchange of the indexes 1
and 2. Thus, associated with the zone for (A,B), there is a partner-zone for
(B,A), where the A- and B-atoms interchange their role.

\subsection{Immiscible phase}

For $[A-B]$ , $u$ and $v$ are distributed in three domains ($0,r_{b}$), ($%
r_{b},r_{c}$), and ($r_{c},\sqrt{2\varepsilon _{2}^{\prime }}$), where
\begin{equation}
r_{c}^{5}=\frac{15\alpha _{1}\alpha _{2}}{2\mathbf{\alpha }Y_{1}}(1+\frac{%
\beta _{1}}{\alpha _{1}}(1-\mathfrak{C}))  \label{eq19}
\end{equation}%
\begin{equation}
r_{b}^{5}=-\frac{15\alpha _{1}\alpha _{2}}{2\mathbf{\alpha }Y_{2}}(1+\frac{%
\beta _{2}}{\alpha _{2}}-\mathfrak{C})  \label{eq20}
\end{equation}

and $\mathfrak{C}$ satisfies the following equation
\begin{equation}
15\alpha _{2}\mathfrak{C}=(2\beta _{2}Y_{1}r_{c}^{2}+2\alpha
_{2}Y_{2}r_{b}^{2})^{5/2}  \label{eq21}
\end{equation}

When this equation has been solved $\mathfrak{C}$ can be obtained. If $%
\mathfrak{C}$ ensure that the right sides of eqs.(\ref{eq19},\ref{eq20}) are
positive, we can obtain $r_{c}$, $r_{b}$.

Then, in the first domain ($0,r_{b}$)
\begin{eqnarray}
u^{2}/r^{2}=\frac{1}{\alpha _{1}}(\varepsilon _{1}^{\prime }-r^{2}/2) \\
v^{2}/r^{2}=0  \label{eq22}
\end{eqnarray}
where
\begin{equation}
\varepsilon _{1}^{\prime }=\alpha _{1}r_{c}^{2}Y_{1}+\beta _{1}r_{b}^{2}Y_{2}
\label{eq23}
\end{equation}

In the second domain ($r_{b},r_{c}$)
\begin{eqnarray}
u^{2}/r^{2} &=&Y_{1}(r_{c}^{2}-r^{2}) \\
v^{2}/r^{2} &=&Y_{2}(r_{b}^{2}-r^{2})  \label{eq24}
\end{eqnarray}%
In the third domain ($r_{c},\sqrt{2\varepsilon _{2}^{\prime }}$)
\begin{eqnarray}
u^{2}/r^{2} &=&0 \\
v^{2}/r^{2} &=&\frac{1}{\alpha _{2}}(\varepsilon _{2}^{\prime }-r^{2}/2)
\label{eq25}
\end{eqnarray}%
where
\begin{equation}
\varepsilon _{2}^{\prime }=\frac{1}{2}(15\alpha _{2}\mathfrak{C)}^{2/5}.
\label{eq26}
\end{equation}

Since $u^{2}/r^{2}$ and $v^{2}/r^{2}$ should be positive in the second
domain, $Y_{1}>0$ and $Y_{2}<0$ are both necessary. In particular, the
latter condition $Y_{2}<0$ assures the emergence of the shell formed by the
B-atoms (because a negative $Y_{2}$\ assures that $v/r$\ increases with $r$\
as shown in eq.(\ref{eq24})). Note that, for attractive $V_{AB}$, we know
from the definitions of $Y_{1}$ and $Y_{2}$ that they must have the same
sign. Thus the immiscible phase does not exist when the interspecies
interaction is attractive (refer to Fig.1 and 5). Therefore, $C_{AB}$ is
considered as positive in the follows.

Obviously, $0\leq r_{b}\leq r_{c}\leq \sqrt{2\varepsilon _{2}^{\prime }}$ is
further required to assure that the solution is meaningful. When $\mathbf{%
\alpha >0}$\ \ and with the preset $Y_{1}>0$ and $Y_{2}<0$ the condition $%
r_{b}\geq 0$ can be satisfied if $\mathfrak{C}\leq 1+\frac{\beta _{2}}{%
\alpha _{2}}$. Besides, from eqs.(\ref{eq19},\ref{eq20}) we have

\begin{equation}
r_{c}^{5}-r_{b}^{5}=\frac{15\alpha _{2}\mathbf{\alpha }}{(\alpha _{1}-\beta
_{2})(\alpha _{2}-\beta _{1})}(\frac{\alpha _{1}}{\alpha _{2}}+1-\mathfrak{C}%
)  \label{eq27}
\end{equation}

Thus the condition $r_{c}\geq r_{b}$ can be satisfied if $\mathfrak{C}\geq 1+%
\frac{\alpha _{1}}{\alpha _{2}}$. Furthermore, from eq.(\ref{eq26}) and
making use of eq.(\ref{eq21}), we have $2\varepsilon _{2}^{\prime }=2\beta
_{2}Y_{1}r_{c}^{2}+2\alpha _{2}Y_{2}r_{b}^{2}$.\ Note that, from eq.(\ref{6}%
) we have
\begin{equation}
\qquad \qquad 2\alpha _{2}Y_{2}+2\beta _{2}Y_{1}=1  \label{eq28}
\end{equation}

Then, $2\varepsilon _{2}^{\prime }=r_{c}^{2}-2\alpha
_{2}Y_{2}(r_{c}^{2}-r_{b}^{2})$. Thus, when $Y_{2}$\ is negative and $%
r_{c}\geq r_{b}$ holds, the condition $\sqrt{2\varepsilon _{2}^{\prime }}%
\geq r_{c}$ is satisfied. To conclude, if we can find out a $\mathfrak{C}$
satisfies eq.(\ref{eq21}) and is lying in the range \ $1+\frac{\alpha _{1}}{%
\alpha _{2}}\leq \mathfrak{C\leq }1+\frac{\beta _{2}}{\alpha _{2}}$, then
the $u$ and $v$\ given above is a physically meaningful solution. By solving
eq.(\ref{eq21}) numerically, a unique $\mathfrak{C}$\ ranging inside the
above range

can be found. Thus, we obtain the solution for the immiscible phase.

Let us find out the zone in the parameter-space that supports the $[A-B]$
solution. When $\mathfrak{C=}1+\frac{\beta _{2}}{\alpha _{2}}$, we have $%
r_{b}=0$, $r_{c}=\sqrt{X_{1}/Y_{1}}\equiv r_{a}$, and $\varepsilon
_{2}^{\prime }=\varepsilon _{2}$.\ Thus the wave functions of the $[A-B]$
and those of (A,B) with $X_{2}=0$ overlap. Recall that $X_{2}=0$\ is a
border of (A,B). Thus the zone of $[A-B]$ is adjacent to the zone of (A,B),
and they have the common border $X_{2}=0$.

When $C_{AB}\rightarrow C_{0}$ (or $\mathbf{\alpha \rightarrow 0}$), we know
from eq.(\ref{eq27}) that $r_{b}\rightarrow r_{c}$, thus the shell tends to
separate from the core completely. Examples are shown in Fig.3. Since $%
r_{b}>r_{c}$ is not allowed, $\mathbf{\alpha =0}$ should be the other
border. Note that, when $\mathbf{\alpha =0}$, the CGP has no physical
solution. Thus, a complete separation between the shell and the core is not
allowed. To conclude, the zone of $[A-B]$ requires $Y_{1}>0$ and $Y_{2}<0$
and is bound by the borders $X_{2}=0$ and $\mathbf{\alpha =0}$.

The case with $\mathbf{\alpha <0}$\ is associated with metastable states,
they can be similarly discussed, if necessary. The partner zone of $[A-B]$ ,
namely, $[B-A]$ , can be similarly obtained by interchanging $u$ and $v$,
and interchanging the indexes 1 and 2.

\end{document}